\documentclass[english,tightenlines,eqsecnum,amsmath,amssymb,aps,prd,superscriptaddress,nofootinbib,showpacs,floats]{article}
\usepackage{epsfig}
\usepackage{babel}
\usepackage{authblk}
\usepackage{amsmath,amsfonts,amssymb,multicol}
\usepackage[usenames]{xcolor}
\definecolor{AHZ}{rgb}{0.0,0.0,0.9}
\definecolor{AHZ1}{rgb}{1,0.0,0.1}
\usepackage[colorlinks,linkcolor=AHZ,citecolor=AHZ1,bookmarks]{hyperref}
\long\def\/*#1*/{}
\usepackage{color}

{\definecolor{RED}{rgb}{1,0,0}
\definecolor{GREEN}{rgb}{0,1,0}
\definecolor{BLUE}{rgb}{0,0,1}
\usepackage{graphicx,caption,subcaption}

\begin{document}

\title{Interpretation of $f({\sf R},{\sf T})$ gravity in terms of a conserved effective fluid}
\author[1]{Hamid Shabani,\thanks{h.shabani@phys.usb.ac.ir}}
\author[2]{Amir Hadi Ziaie,\thanks{ah.ziaie@gmail.com}}

\affil[1]{Physics Department, Faculty of Sciences, University of Sistan and Baluchestan, Zahedan, Iran}
\affil[2]{Department of Physics, Kahnooj Branch, Islamic Azad University, Kerman, Iran}
\date{\today}
%
\maketitle
\begin{abstract}
\noindent
In the present work we introduce a novel approach to study $f({\sf R},{\sf T})$ gravity theory from a different perspective. Here, ${\sf T}$ denotes the trace of energy-momentum tensor ({\sf EMT}) of matter fluids. The usual method (as discussed in the literature) is to choose an $h({\sf T})$ function and then solve for the resulted Friedman equations. Nevertheless, our aim here is, without loss of generality, to reformulate a particular class of $f({\sf R},{\sf T})$ gravity models in which the Einstein-Hilbert action is promoted by an arbitrary function of the trace of {\sf EMT}. The strategy is the redefinition of the equation of motion in terms of the components of an effective fluid. We show that in this case the {\sf EMT} is automatically conserved. As we shall see, adopting such a point of view (at least) in $f({\sf R},{\sf T})$ gravity is accompanied by two significant points. On one hand, $h({\sf T})$ function is chosen based upon a physical concept and on the other, we clearly understand the overall or effective behavior of matter in terms of a conserved effective fluid. To illustrate the idea, we study some models in which different physical properties for the effective fluid is attributed to each model. Particularly, we discuss models with constant effective density, constant effective pressure and constant effective equation of state ({\sf EoS}) parameter. Moreover, two models with a relation between the effective density and the effective pressure will be considered. An elegant result is that in $f({\sf R},{\sf T})$ gravity, there is a possibility that a perfect fluid could effectively behave as a modified Chaplygin gas with four free parameters.
\end{abstract}
\section{Introduction}\label{intro}
Einstein\rq{}s Theory of General Relativity ({\sf GR}) is a marvelous accomplishment and has been confirmed as a successful theory over many years of experimental tests \cite{ExGrdata}. Despite the successful outcomes achieved by {\sf GR} in describing the Universe and the Solar System, it is a well accepted idea that {\sf GR} along with cosmological constant is not the ultimate gravitation theory, but an extremely good approximation valid within the present day range of observational measurements. Recently, modified gravity theories ({\sf MGT}) have been widely investigated with the hope to find observationally consistent alternatives for {\sf GR}. This is due to the recently observational data~\cite{Riess1999,Abazajian2005,Hinshaw2013,Ade2016} which have shown that there may exist two mysterious essences that could possibly impress the evolution of the Universe. In this sense, observations have led to the introduction of additional ad-hoc concepts like Dark Energy ({\sf DE}), which results in accelerated expansion of the Universe, and Dark Matter ({\sf DM}), which is responsible for the formation of galaxy clusters, within the standard model of cosmology. However the ambiguous nature of these components may be interpreted as the possibility of failure of {\sf GR} on large (infrared regimes) scales. The $\Lambda${\sf CDM} or the concordance model~\cite{Ostriker},  may be the simplest cosmological model that incorporates these two dark components. In this scenario, a cosmological constant $\Lambda$ is added to the usual Einstein-Hilbert action. However, the cosmological constant suffers from a fine-tuning problem related to its energy scale, if one attributes it to vacuum energy~\cite{Weinberg1989,Padmanabhan2013}. This problem has led to encouraging motivations to search for alternative models of {\sf DE} beyond the $\Lambda${\sf CDM} model.

There are two main approaches to deal with such issues; one can involve different ingredients (such as scalar fields, vector fields or other forms of matter fields) within the {\sf GR} action and then study the possible effects that could arise as the outcome. As an alternative way, one may alter the background theory (which may be supported by a fundamental idea, such as Brans-Dicke~\cite{Faraoni2004} theory) and analyze subsequent equations of motion in order to find new features (which may or may not be compatible with the astronomical observations). One may also use both approach simultaneously, i,e, including a new ingredient as the matter content together with modifying the background geometry~\cite{Baffou2015,Shabani2016,Baffou2017}.
It is however possible that applying both approaches to the same model does not give the same results. For example, there still are debates about the correspondence of $f({\sf R})$ gravities with scalar-tensor theories~\cite{de Felice2010,Bahamonde2017}. After designing a theory what remains is to check its consistency with observational data, that is, seeking for physical validity of the theory. In the case of gravitational theories, it is important to check whether or not they are cosmologically viable\footnote{One can refer to~\cite{Will1993} for more details.}.

In the present work we consider cosmological behavior of $f({\sf R},{\sf T})$ gravity theories~\cite{Harko2011} from different points of view. In the $f({\sf R},{\sf T})$ gravity proposal which its aspects are still under consideration within different scenarios~\cite{Alvarenga2013,Shabani2013,Harko2014,Shabani2014,Sharif2014,Alves2016,
Sun2016,odzasehash,Fayaz2016,Zaregonbadi2016-1,Zaregonbadi2016-2,Moraes2016,Shabani2017-1,Shabani2017-2,grwavefrt}, one uses the trace of {\sf EMT} coupled to the Ricci curvature scalar ${\sf R}$ in an arbitrary form. The aim is to construct models in which gravitational and cosmological effects emanate from an unusual interactions between matter and geometrical sectors or even from the effects of some unknown matter fields. A major problem about these type of theories (which up to now, are less questioned in the literature) is that how matter interacts with geometry; or equivalently, can we describe the effects of unusual contribution of matter within the field equations in an unambiguous way? One may also translate these questions into saying that, what is the effective/overall behavior of matter contribution within $f({\sf R},{\sf T})$ gravity? Another problem which is implicit in these theories, is that how one can choose a physically meaningful $f({\sf R},{\sf T})$ function. If there exists a logical and physically reliable way, then a more profound understanding can be achieved from the study of these theories. Also, it may be more helpful to construct observationally viable models.

To demonstrate these two questions (and find a better understanding of the physical concepts lying behind the theory) we reformulate cosmological field equations of minimally coupled form of $f({\sf R},{\sf T})$ gravity, i.e, $f({\sf R},{\sf T})={\sf R}+\alpha\kappa^{2}h({\sf T})$, where $\kappa^{2}\equiv8\pi G/c^4$ is the gravitational coupling constant. We define an effective fluid whose components are defined by an actual perfect fluid with {\sf EoS} parameter $w$. Thus, briefly speaking, we leave the common view on $f({\sf R},{\sf T})={\sf R}+\alpha\kappa^{2}h({\sf T})$ field equation which has the following properties 
\begin{align}\label{intro1}
&\left\{\begin{array}{ll}
\bullet~\mbox{original~field~equations~introduced~in~\cite{Harko2011}},\\
\bullet~\mbox{matter sector involves \lq\lq{}real\rq\rq{}~matter~fields~ e.g.,~a~perfect~fluid~with}~p=w \rho,\\
\bullet~\mbox{no~criterion~for~choosing~a~physically~meaningful~$h$({\sf T})~function,}\\
\bullet~\mbox{obscure~interaction~of~the~real~fluid~with~geometry},                            
\end{array}
\right.
\end{align}
and construct a novel landscape (generalized {\sf GR} field equations) with new features
\begin{align}\label{intro2}
&\left\{\begin{array}{ll}
\star~\mbox{{\sf GR} field equations with a modified source (which is called \lq\lq{}effective\rq\rq{}}\\
~~\mbox{or \lq\lq{}exotic\rq\rq{} fluid),}\\
\star~\mbox{this effective fluid may generally behave completely different from}\\
~~\mbox{the real matter field (various possibilities will be discussed in this work),}\\
\star~\mbox{one~can~obtain~$h$({\sf T})~for~an~effective~fluid~with~a~specified~{\sf EoS},}\\
\star~\mbox{usual gravitational interaction of the effective fluid through}\\
~~\mbox{the {\sf GR} fields equation.}               
\end{array}
\right.
\end{align}
Therefore, we will show that, $f({\sf R},{\sf T})$ gravity field equations with a real fluid such as a perfect fluid can be translated into the {\sf GR} with an exotic fluid with $p_{\rm{(eff)}}(\sf{T})$ and $\rho_{\rm{(eff)}}(\sf{T})$ like the modified Chaplygin gas~\cite{Benaoum02} (as we shall obtain such a description in the upcoming sections).
 
Such a picture is not unreasonable. Studies of the energy conditions in $f({\sf R},{\sf T})$ gravity may highlight the importance of the effective picture. In~\cite{Sharif13}, the authors have shown that the energy conditions in $f({\sf R},{\sf T})$ gravity can be written in the same way as those given in GR but for an effective fluid. Thus, it seems natural to define a new framework (i.e., the \lq\lq{}{\it effective picture}\rq\rq{}) and interpret the resulted field equations through the effective elements. By further studying $f({\sf R},{\sf T})$ gravity in the framework of effective picture, one finds some coincidence with other models. To elaborate this fact we briefly point out two examples.

In the present work we discuss different classes of models characterized by different effective {\sf EoS}s or equivalently different $h({\sf T})$ functions. It will be shown that the effective quantities can be generally written as
\begin{align}
&\rho_{({\sf eff})}({\sf T})=\beta_{1}{\sf T}+\beta_{2}{\sf T}^{\gamma}+\tilde{\rho}_{({\sf eff})}\label{intro3},\\
&p_{({\sf eff})}({\sf T})=\lambda_{1}{\sf T}+\lambda_{2}{\sf T}^{\gamma}+\tilde{p}_{({\sf eff})}\label{intro4},
\end{align}
where, $\beta_{i},~\lambda_{i},~\gamma,~\tilde{\rho}_{({\sf eff})}$ and $\tilde{p}_{({\sf eff})}$ are constants. Eliminating the trace between the expressions given for effective density and pressure we obtain an {\sf EoS} as follows
\begin{align}
\frac{\lambda _1 \left(\rho -\tilde{\rho}\right)-\beta _1 \left(p-\tilde{p}\right)}{\beta _2 \lambda _1-\beta _1 \lambda _2}=\left[\frac{\lambda _2 \left(\rho -\tilde{\rho}\right)-\beta _2 \left(p-\tilde{p}\right)}{\beta _1 \lambda _2-\beta _2 \lambda _1}\right]^{\gamma }\label{intro5},
\end{align}
where the subscript \lq\lq{}{\sf eff}\rq\rq{} has been dropped. Hence, the unusual gravitational interaction hidden in $f({\sf R,T})$ gravity may be effectively translated as usual gravitational interaction of the spacetime curvature with an exotic fluid which obeys the {\sf EoS} of the form (\ref{intro5}). It is interesting to note that a reduced form of (\ref{intro5}) has been already introduced in~\cite{Stefancic05,Nojiri05} in order to investigate the cosmological consequences of models with mixture of quintessence and phantom matter fields. One can also find related works in the literature where an \lq\lq{}exotic fluid\rq\rq{} is introduced to solve some unanswered questions in cosmology. All this exotic fluids obey different subclasses of the {\sf EoS} given in (\ref{intro5}) and the underlying models may be called models with \lq\lq{}modified equation of state\rq\rq{} ({\sf MEoS}). 

For example we address some relevant works; an {\sf EoS} of the form $p=-\rho+\gamma \rho^{\lambda}$ has been employed in~\cite{Barrow90}, in order to obtain power-law and exponential inflationary solutions. The particular case of $\lambda=1/2$ has been discussed in~\cite{Stefancic052,Nojiri052,Nojiri05} to study the expansion of the Universe. An exotic component with $p=A\rho-B \rho^{1/2}$ has been investigated in~\cite{Mukherjee06} in order to study the scenario of emergent Universe. In~\cite{Contreras16} the same {\sf EoS} is applied with $A=-1$. The authors of~\cite{Babichev05} have considered different cosmological aspects of a more simple form, i.e.,  $p_{{\sf DE}}=\alpha(\rho_{{\sf DE}}-\rho_{0})$ and in~\cite{Contreras17} cosmological bouncing solutions have been considered.

Therefore, in view of the above considerations, a bridge from $f({\sf R},{\sf T})$ gravity within the context of the effective picture to {\sf MEoS} theories can be found. In these two types of theories, there is an exoticness which in the former is hidden behind the gravitational interactions and in the former shows itself as a mysterious fluid. However, in $f({\sf R},{\sf T})$ gravity in effective picture, the action is determined unlike the {\sf MEoS} theories.

As another coincidence, in the Rastall theory~\cite{Rastal72} the usual conservation of the {\sf EMT} is modified as ${\sf T}^{\alpha}_{\,\,\beta;\alpha}=\lambda {\sf R}_{,\beta}$ (where $;$ and $,$ denote covariant and partial differentiations and $\lambda$ is a constant). The field equations can then be obtained as follows
\begin{align}\label{intro6}
{\sf G}_{\mu\nu}+k\,\lambda\,{\sf g}_{\mu\nu} {\sf R}=k {\sf T}_{\mu\nu}.
\end{align}
Using the trace equation one obtains ${\sf R}=k{\sf T}/(4k\lambda-1)\equiv \tilde{k}{\sf T}$. Thus, the Rastall field equation can be recast into the following form
\begin{align}\label{intro7}
{\sf G}_{\mu\nu}={\sf T}_{\mu\nu}-\tilde{k}\,k\,\lambda\,{\sf g}_{\mu\nu}{\sf T}\equiv {\sf T}^{(\rm{eff})}_{\mu\nu}.
\end{align}
As can be seen, the rastall theory may be categorized as a particular class of $f({\sf R},{\sf T})$ gravity in the effective picture. In the original paper~\cite{Rastal72}, there has been also discussed that this theory is equivalent to {\sf GR} with an ideal fluid with new pressure and density (shown by $p'$ and $\rho'$), and thus different {\sf EoS} in comparison to the {\sf EoS} governing the {\sf EMT} components at the right hand side of equation (\ref{intro6}). Therefore, one can find the footprint of the effective representation of $f({\sf R},{\sf T})$ in various works. As an application of such a viewpoint we have investigated non-singular cosmological solutions in $f({\sf R},{\sf T})$ gravity in~\cite{Shabani2017-4}.


The paper is planned as follows: In Section~\ref{eff1}, we review the field equations of $f({\sf R},{\sf T})$ gravity, very briefly. In Section~\ref{eff2}, we introduce an effective fluid and obtain the equation governing the conservation of {\sf EMT} for a perfect fluid. Section~\ref{eff3} is devoted to discuss some cosmological models based on the idea of the effective fluid. Finally, in~\ref{con} we give the summery and conclusion.
%
\section{Field equations of $f({\sf R},{\sf T})$ gravity}\label{eff1}
In this section, we concisely review the field equations of $f({\sf R},{\sf T})$ modified gravity and discuss the effects of 
violation of {\sf EMT}. The action of $f({\sf R},{\sf T})$ gravity has been initially introduced as~\cite{Harko2011}
\begin{align}\label{action}
S=\int \sqrt{-g} d^{4} x \left[\frac{1}{2\kappa^{2}} f\Big{(}{\sf R}, {\sf T}\Big{)}+{\sf L}^{\rm{(m)}}\right],
\end{align}
where ${\sf R}$, ${\sf T}\equiv g^{\mu \nu} {\sf T}_{\mu \nu}$, ${\sf L}^{\rm{(m)}}$ are the Ricci
curvature scalar, the trace of {\sf EMT} and the Lagrangian of 
pressure-less matter, respectively. The determinant of the metric is denoted by $g$, 
and we have set the units so that $c=1$. The {\sf EMT} for matter fields is defined as follows
\begin{align}\label{Euler-Lagrange}
{\sf T}_{\mu \nu}\equiv-\frac{2}{\sqrt{-g}}
\frac{\delta\left[\sqrt{-g}{\sf L}^{\rm{(m)}}\right]}{\delta g^{\mu \nu}}.
\end{align}
Varying action (\ref{action}) with respect to metric leads to the following field equation~\cite{Harko2011}
\begin{align}\label{fRT field equations}
&F({\sf R},{\sf T}) {\sf R}_{\mu \nu}-\frac{1}{2} f({\sf R},{\sf T}) g_{\mu \nu}+\Big{(} g_{\mu \nu}
\square -\triangledown_{\mu} \triangledown_{\nu}\Big{)}F({\sf R},{\sf T})=\nonumber\\
&\Big{(}\kappa^{2}-{\mathcal F}({\sf R},{\sf T})\Big{)}T_{\mu \nu}-\mathcal {F}({\sf R},{\sf T})\mathbf
{\Theta_{\mu \nu}},
\end{align}
where
\begin{align}\label{theta}
\mathbf{\Theta_{\mu \nu}}\equiv g^{\alpha \beta}\frac{\delta
{\sf T}_{\alpha \beta}}{\delta g^{\mu \nu}}=-2{\sf T}_{\alpha \beta}+g_{\alpha \beta}{\sf L}^{\rm{(m)}}-2g^{\alpha \beta}\frac{\partial^{2}{\sf L}^{\rm{(m)}}}{\partial g^{\alpha \beta}\partial g^{\mu \nu} },
\end{align}
and, the following definitions has been used for the sake of convenience
\begin{align}\label{f definitions1}
{\mathcal F}({\sf R},{\sf T}) \equiv \frac{\partial f({\sf R},{\sf T})}{\partial {\sf T}}~~~~~
~~~~~\mbox{and}~~~~~~~~~~
F({\sf R},{\sf T}) \equiv \frac{\partial f({\sf R},{\sf T})}{\partial {\sf R}}.
\end{align}
Applying the spatially flat Friedmann--Lema\^{\i}tre--Robertson--Walker ({\sf FLRW}) metric
\begin{align}\label{metricFRW}
ds^{2}=-dt^{2}+a^{2}(t) \Big{(}dr^{2}+r^{2}d\Omega^2\Big{)},
\end{align}
to field equation (\ref{fRT field equations}) for a perfect fluid gives
\begin{align}\label{first}
&3H^{2}F({\sf R},{\sf T})+\frac{1}{2} \Big{(}f({\sf R},{\sf T})-F({\sf R},{\sf T}){\sf R}\Big{)}+3\dot{F}({\sf R},{\sf T})H=\nonumber\\
&\Big{(}\kappa^{2} +{\mathcal F} ({\sf R},{\sf T})\Big{)}\rho+{\mathcal F} ({\sf R},{\sf T})p,
\end{align}
as the modified Friedmann equation, and
\begin{align}\label{second}
&2F({\sf R},{\sf T}) \dot{H}+\ddot{F} ({\sf R},{\sf T})-\dot{F} ({\sf R},{\sf T}) H=-\Big{(}\kappa^{2}+{\mathcal F} ({\sf R},{\sf T})\Big{)}(\rho+p),
\end{align}
as the modified Raychaudhuri equation. In expression (\ref{theta}) we have 
used ${\sf L}^{\rm{(m)}}=p$ for a perfect fluid. Also, in metric (\ref{metricFRW}), $a$ denotes the scale 
factor of the Universe and in equations (\ref{first}) and (\ref{second}), $H$ indicates the 
Hubble parameter. For the field equation (\ref{fRT field equations}), the Bianchi identity  
leads the following covariant equation
\begin{align}\label{relation}
&(\kappa^{2} +\mathcal {F})\nabla^{\mu}{\sf T}_{\mu \nu}+\frac{1}{2}\mathcal {F}\nabla_{\mu}{\sf T}
+{\sf T}_{\mu \nu}\nabla^{\mu}\mathcal {F}-\nabla_{\nu}(p\mathcal{F})=0,
\end{align}
where the argument of $\mathcal {F}({\sf R},{\sf T})$ has been dropped for abbreviation.
As it is seen from (\ref{relation}) and also it has recently discussed in several works, the {\sf EMT} is not automatically conserved in $f({\sf R},{\sf T})$ gravity. In ~\cite{Harko2014} it has been shown that from thermodynamic point of view this non-conservation is equivalent to an irreversible matter creation processes which fundamental particle physics could justify it. Such a particle creation corresponds to a direction of energy flow from the gravitational field to the created matter constituents. In the next section, we present a novel feature of $f({\sf R},{\sf T})$ gravity which is the consequence of violation of the {\sf EMT} conservation of normal matter. We shall see that equation (\ref{relation}) actually plays a key role in determining the different aspects of this theory.
\section{$f({\sf R},{\sf T})$ gravity in terms of a conserved effective fluid}\label{eff2}
 In this section, $f({\sf R},{\sf T})$ modified gravity will be investigated from a new and different window. Though the {\sf EMT} is not automatically conserved in $f({\sf R},{\sf T})$ gravity, we are still capable of defining an ``effective"  conserved fluid. The study of $f({\sf R},{\sf T})$ gravity theories based on such a fluid, gives deeper insight into the role of the trace of {\sf EMT} in these theories. We include a single barotropic perfect fluid with {\sf EoS} $p=w\rho$. Hereafter, without loss of generality it is focused on a Lagrangian with minimal couplings between the Ricci scalar and the trace of {\sf EMT} which is given as
\begin{align}\label{minimal}
f({\sf R},{\sf T})={\sf R}+\alpha\kappa^{2} h({\sf T}).
\end{align}
For this class of $f({\sf R},{\sf T})$ models field equations (\ref{first}) and (\ref{second}) become
\begin{align}\label{eom1}
3H^{2}=\kappa^{2}\Bigg{\{}\Big{[}1+(1+w)\alpha h'\Big{]}\frac{{\sf T}}{3w-1}-\frac{\alpha h}{2}\Bigg{\}},
\end{align}
and
\begin{align}\label{eom2}
2\dot{H}=-\kappa^{2}\frac{w+1}{3w-1}(1+\alpha h'){\sf T}.
\end{align}
Also, substituting the choice (\ref{minimal}) in equation (\ref{relation}) leads to 
\begin{align}\label{relation-w}
\Big{(}1 + \frac{\alpha}{2}(3-w)h' + \alpha (1+w) Th''\Big{)}\dot{{\sf T}}+
3H(1+w)\Big{(}1 + \alpha h'\Big{)}{\sf T}=0.
\end{align}
A straightforward technique can be, solving equation (\ref{relation-w}) using a specific $h({\sf T})$ function. This gives ${\sf T}$ in terms of the scale factor $a$. Nevertheless, we adopt a different approach. To this end we rewrite equation (\ref{eom1}) as follows
\begin{align}\label{re1}
3H^{2}=\kappa^{2}\rho_{(\rm{eff})}({\sf T}),
\end{align}
where,
\begin{align}\label{re2}
\rho_{(\rm{eff})}({\sf T})\equiv \Big{[}1+(1+w)\alpha h'\Big{]}\frac{{\sf T}}{3w-1}-\frac{\alpha h}{2}.
\end{align}
With this definition, equation (\ref{eom2}) can be rewritten as
\begin{align}\label{re4}
2\dot{H}=-\kappa^{2}\Big{[}\rho_{(\rm{eff})}({\sf T})+\frac{w}{3w-1}{\sf T}+\frac{\alpha h}{2}\Big{]}.
\end{align}
The acceleration of expansion of the Universe can be obtained from combining (\ref{re1}) and (\ref{re4}), which gives
\begin{align}\label{re5}
\frac{\ddot{a}}{a}=-\frac{\kappa^{2}}{6}\Big{(}\rho_{(\rm{eff})}({\sf T})+3p_{(\rm{eff})}({\sf T})\Big{)},
\end{align}
where, the effective pressure is defined as 
\begin{align}\label{re6}
p_{(\rm{eff})}({\sf T})\equiv\frac{w}{3w-1}{\sf T}+\frac{\alpha h}{2}.
\end{align}
Definitions (\ref{re2}) and (\ref{re6}) give the effective density and pressure in terms of the trace ${\sf T}$. Once ${\sf T}$ and $h({\sf T})$ are determined in terms of the scale factor, the effective density and pressure would be explicitly specified. There is a crucial step to complete the above method which leads to determination of ${\sf T}$ for a given $h({\sf T})$ function. One can rewrite equation (\ref{relation-w}) in terms of $\rho_{(\rm{eff})}$ and $p_{(\rm{eff})}$ as follows
\begin{align}\label{re7}
\dot{\rho}_{(\rm{eff})}+3H(\rho_{(\rm{eff})}+p_{(\rm{eff})})=0,
\end{align}
where the arguments are dropped for simplicity. Therefore, one can reformulate minimally coupled forms of $f({\sf R},{\sf T})$ gravity
in terms of a conserved effective fluid with a defined effective density and pressure. For this effective fluid one can define an effective {\sf EoS} parameter as follows 
\begin{align}\label{re8}
\mathcal{W}_{(\rm{eff})}\equiv \frac{p_{(\rm{eff})}}{\rho_{(\rm{eff})}}=-1+\frac{2(1+w)(1+\alpha h'){\sf T}}{2\left[1+(1+w)\alpha h'\right]{\sf T}-(3w-1)\alpha h}.
\end{align}
The above picture will be completed only when $h({\sf T})$ function is specified. But the crucial  step is how to choose a suitable $h({\sf T})$ function. In the original formulation of $f({\sf R},{\sf T})$ gravity there is no explicit benchmark to help one to choose a particular 
$h({\sf T})$ function which leads to a remarkable physics. All things that one can do is to apply a function $h({\sf T})$ without any prior intuition. However, in the above picture one can obtain the $h({\sf T})$ function which is related to a particular property of the effective fluid. That is, in the language of effective fluid there is an elucidated procedure. In the next section we consider behavior of an actual perfect fluid in terms of an effective fluid. That is, we investigate how a perfect fluid with the {\sf EoS} parameter $w$, behaves as a mysterious cosmic fluid with $p_{(\rm{eff})}$ and $\rho_{(\rm{eff})}$. 
\section{Effective behavior of matter in $f({\sf R},{\sf T})$ gravity}\label{eff3}
In this section, we consider some particular cosmological models in $f({\sf R},{\sf T})$ gravity which can be described by an effective cosmic fluid. Such an artificial fluid is a translation of the interactions of the normal (perfect) fluid with the geometry, which is interpreted as an effective behavior of a single (exotic) fluid within the framework of $f({\sf R},{\sf T})$ gravity. We then continue our investigations as follows; firstly, some effective fluids will be specified by considering a property of or a relation between their pressure or density profiles. We then try to obtain the $h({\sf T})$ function (in terms of the scale factor) which corresponds to the specific property that the effective fluid could have. Therefore, the effective fluid can be determined. In the last step, one may analyze cosmological consequences of such an effective fluid.
\subsection{Effective fluid with constant density, $\rho_{(\rm{eff})}=\rho_{(\rm{eff}),c}$}\label{sub1}
As the first case, we discuss the class of $f({\sf R},{\sf T})$ gravity models which have a constant effective density.  For these type of models from conservation equation (\ref{re7}) we conclude that $p_{(\rm{eff})}=-\rho_{(\rm{eff}),c}$, where $\rho_{(\rm{eff}),c}$ is a constant; therefore from definitions (\ref{re2}) and (\ref{re6}) we obtain two equations for $h({\sf T})$, as follows
\begin{align}\label{eff3-1}
\Big{[}1+(1+w)\alpha h'\Big{]}\frac{{\sf T}}{3w-1}-\frac{\alpha h}{2}=\rho_{(\rm{eff}),c},
\end{align}
and
\begin{align}\label{eff3-2}
\frac{w}{3w-1}{\sf T}+\frac{\alpha h}{2}=-\rho_{(\rm{eff}),c}.
\end{align}
One can check that equations (\ref{eff3-1}) and (\ref{eff3-2}) accept the same solution only for two particular cases. These solutions are obtained as
\begin{align}
&h_{\rho_{(\rm{eff}),c}}^{w=1}({\sf T})=-\frac{{\sf T}+2 \rho _{(\rm{eff}),c}}{\alpha }~~~\mbox{for}~~~w=1,\label{eff3-3}\\
&h_{\rho_{(\rm{eff}),c}}^{w=-1}({\sf T})=-\frac{{\sf T}+4 \rho_{(\rm{eff}),c}}{2 \alpha }~~~\mbox{for}~~~w=-1.\label{eff3-43}
\end{align}
Therefore, a perfect fluid with either of the {\sf EoS} parameters $w=+1$ or $w=-1$ leads to a de Sitter state at the late times, provided that functions (\ref{eff3-3}) or (\ref{eff3-43}) are chosen, respectively. In these cases the conditions $H=constant$ and $\dot{H}=0$ hold as can be seen from (\ref{re1}) and (\ref{re4}). In terms of the effective fluid properties, these two cases yield $\mathcal{W}_{(\rm{eff})}=-1$. Moreover, from equation (\ref{re1}) we may intuitively imagine that the origin of {\sf DE} can be rooted in the interaction between ordinary matter fields and geometry so that such an interaction could effectively produces a late time accelerated expansion regime.
\subsection{Effective fluid with constant pressure, $p_{(\rm{eff})}=p_{(\rm{eff}),c}$}\label{sub2}
\begin{figure}[h!]
\centering
\centerline{\epsfig{figure=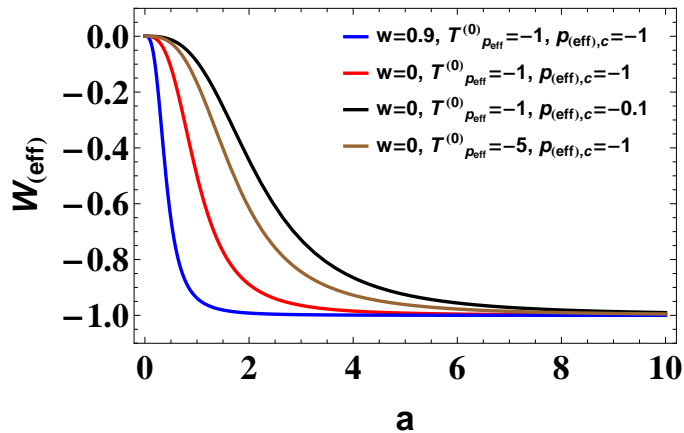,width=9cm}}
\caption{The behavior of effective {\sf EoS} for the models with constant effective pressure.}
\label{fig1}
\end{figure}

The case of an effective fluid with the property that $p_{(\rm{eff})}=constant=p_{(\rm{eff}),c}$, leads to the following expression for $h({\sf T})$ function
\begin{align}\label{eff3-4}
h_{p_{(\rm{eff}),c}}^{w}({\sf T})=\frac{2}{\alpha}\left(\frac{w{\sf T}}{(1 -3w)}+p_{(\rm{eff}),c}\right).
\end{align}
Substituting solution (\ref{eff3-4}) into equation (\ref{relation-w}) gives the trace of {\sf EMT} in terms of the scale factor, as
\begin{align}\label{eff3-5}
{\sf T}_{p_{(\rm{eff}),c}}^{w}(a)={\sf T}^{(0)}_{p_{(\rm{eff})}} a^{-3},
\end{align}
where ${\sf T}^{(0)}_{p_{(\rm{eff})}}$ is an integration constant which can be set according to its present value. The effective density is obtained as
\begin{align}\label{eff3-55}
\rho_{p_{(\rm{eff}),c}}({\sf T})=\frac{\left(w^2-1\right)}{(1-3 w)^2}{\sf T}-p_{(\rm{eff}),c},
\end{align}
Therefore, for the effective {\sf EoS} parameter, $\mathcal{W}_{{(\rm{eff})}}$ we have
\begin{align}\label{eff3-6}
\mathcal{W}_{{(\rm{eff})}}=-1+\left[1-\frac{(1-3 w)^{2}p_{(\rm{eff}),c}}{\left(w^2-1\right){\sf T}}\right]^{-1}.
\end{align}
The above solution shows that for non-zero values of $p_{(\rm{eff}),c}$ a transition from the matter dominated universe to de Sitter era occurs. From (\ref{eff3-55}) we see that in the case of $w=\pm1$ there is only a de Sitter era at late times. Comparing the effective density given in (\ref{eff3-55}) with the effective density for $\Lambda${\sf CDM} model, we may conclude that the cosmological constant in this case, can be translated as an effective pressure. In $f({\sf R},{\sf T})$ gravity, the effective fluid with zero pressure does not include an accelerated expansion era at late time. In Figure~\ref{fig1} we have plotted the evolution of $\mathcal{W}_{{(\rm{eff})}}$ for different values of $w$, $C_{p_{(\rm{eff}),c}}$ and $p_{(\rm{eff}),c}$ parameters.
All the curves in this figure have been drawn so as to let one to be capable of comparing the effects of each constant. For example, comparing the blue and red curves indicates that the larger the value of $w$ leads the de Sitter epoch to occur sooner. Also, comparing the red and black curves shows that the larger values of effective pressure lead to sooner transition to the de Sitter era. 
\subsection{Effective fluid mimicking a perfect fluid: $p_{(\rm{eff})}=\mathcal{W}_{(\rm{eff})}\rho_{(\rm{eff})}$}\label{sub3}
In subsection~\ref{sub2} we observed that the case of an effective fluid with constant pressure leads to a trace dependent effective {\sf EoS} parameter. Nevertheless, in the view of such an effective behavior, it is also possible that a perfect fluid with {\sf EoS} parameter $w$ behaves as an effective perfect fluid with constant $\mathcal{W}_{\rm{(eff)}}$ parameter. For these cases, the relation $p_{(\rm{eff})}=\mathcal{W}_{(\rm{eff})}\rho_{(\rm{eff})}$ yields to the following differential equation for $h({\sf T})$ function
\begin{align}\label{eff3-7}
-2\alpha\mathcal{W}_{(\rm{eff})}(w+1){\sf T}h'({\sf T}) +\alpha  (3 w-1) \left(\mathcal{W}_{(\rm{eff})}+1\right) h({\sf T})+2(w-\mathcal{W}_{(\rm{eff})}){\sf T}=0,
\end{align}
for which the solution reads
\begin{align}\label{eff3-8}
h^{w}_{\mathcal{W}_{\rm{(eff})}}({\sf T})=\mathcal{C} {\sf T}^{\frac{(3 w-1) \left(\mathcal{W}_{\rm{(eff})}+1\right)}{2 (w+1) \mathcal{W}_{\rm{(eff})}}}+\frac{2 \left(\mathcal{W}_{(\rm{eff})}-w\right)}{\alpha  \left[(w-3) \mathcal{W}_{(\rm{eff})}+3 w-1\right]}{\sf T},
\end{align}
where $\mathcal{C}$ is a constant of integration and a trivial constant is discarded. Substituting solution (\ref{eff3-8}) into equation (\ref{relation-w}) leads to the following differential equation for ${\sf T}(a)$

\begin{align}\label{eff3-9}
\Big{(}\zeta_{1}{\sf T}^{\beta}+\zeta\Big{)}\frac{d{\sf T}}{da}+\frac{3}{a} \left(\mathcal{W}_{\rm{(eff)}}+1\right) \Big{(}\zeta_{2}{\sf T}^{\beta}+\zeta\Big{)}{\sf T}=0,
\end{align}
where
\begin{align}\label{eff3-10}
&\zeta\equiv\frac{2 \left(w^2-1\right)}{(w-3) \mathcal{W}_{\rm{(eff)}}+3 w-1},\nonumber\\
&\zeta_{1}\equiv \frac{\alpha \mathcal{C} (1-3 w)^2 \left(\mathcal{W}_{\rm{(eff)}}+1\right)}{2 (w+1) \mathcal{W}_{\rm{(eff)}}^2}, \nonumber\\
&\zeta_{2}\equiv \frac{\alpha  \mathcal{C} (3 w-1)}{\mathcal{W}_{\rm{(eff)}}}, \nonumber\\
&\beta\equiv\frac{\mathcal{W}_{\rm{(eff)}}(w-3)+3 w-1}{2 \mathcal{W}_{\rm{(eff)}}(w+1)}.
\end{align}
A general solution of equation (\ref{eff3-9}) is of type the inverse of hypergeometric function. However, we can consider more tractable cases. For example, we can check that for a stiff fluid with $w=1$ equation (\ref{eff3-9})
can be simplified so that we readily obtain the solutions as
\begin{align}\label{eff3-11}
&h^{1}_{\mathcal{W}_{\rm{(eff)}}}({\sf T})=\gamma_{\mathcal{W}_{\rm{(eff)}}}^{1}{\sf T}^{\frac{1}{2} \left(\frac{1}{\mathcal{W}_{\rm{(eff)}}}+1\right)}-\frac{{\sf T}}{\alpha },\\
&{\sf T}^{1}_{\mathcal{W}_{\rm{(eff)}}}(a)={\sf T}^{1}_{\mathcal{W}_{\rm{(eff)}},0}a^{-6 \mathcal{W}_{\rm{(eff)}}},\label{eff3-x11}\\
&\rho_{\rm{(eff)},\mathcal{W}_{\rm{(eff)}}}^{1}({\sf T})=\frac{\alpha \gamma_{\mathcal{W}_{\rm{(eff)}}}^{1}}{2 \mathcal{W}_{\rm{(eff)}}} {\sf T}^{\frac{1}{2} \left(\frac{1}{ \mathcal{W}_{\rm{(eff)}}}+1\right)},
\end{align}
where we have labeled functions and constants as $\mathcal{O}^{w}_{\mathcal{W}_{\rm{(eff)}}}({\sf T})$, and $\gamma_{\mathcal{W}_{\rm{(eff)}}}^{w}$ denotes the integration constant of equation (\ref{eff3-7}). Note that, the subscript $0$ in expression (\ref{eff3-x11}) indicates the present values. We therefore observe that a stiff fluid in $f({\sf R},{\sf T})$ gravityو can effectively play the role of a perfect fluid with arbitrary effective {\sf EoS} parameter $\mathcal{W}_{\rm{(eff)}}$. Moreover, an interesting case is when a stiff fluid mimics the behavior of an (effective) ultra relativistic fluid with $\mathcal{W}_{\rm{(eff)}}=1/3$. 

Another examples are cases with $w=0$. However, for these class of models the related equations cannot be solved analytically; we therefore consider two specific cases. A pressure-less fluid can effectively behave as a cosmic string which has the following properties
\begin{align}
&h^{0}_{-1/3}({\sf T})={\sf T} \left( \gamma_{-1/3}^{0}-\frac{\log ({\sf T})}{\alpha }\right)\label{eff3-12}\\
&{\sf T}^{0}_{-1/3}(a)=\exp \left[W\left(\frac{{\sf T}^{0}_{-1/3,0} e^{-\alpha  \gamma_{-1/3}^{0}} \left(\log ({\sf T}^{0}_{-1/3,0})-\alpha   \gamma_{-1/3}^{0}\right)}{a^2}\right)+\alpha  \gamma_{-1/3}^{0}\right],\label{eff3-13}\\
&\rho_{\rm{(eff)},-1/3}^{0}({\sf T})=\frac{3}{2} {\sf T} \left(\log ({\sf T})-\alpha  \gamma_{-1/3}^{0}\right)\label{eff3-14}.
\end{align}
In (\ref{eff3-13}), $W$ is the Lambert-W function which satisfies $d W/dz=W/(W+1) z$. Also, a pressure-less perfect fluid can behave as a stiff fluid with the following properties
\begin{align}
&h^{0}_{1}({\sf T})=\frac{\gamma_{1}^{0}}{{\sf T}}-\frac{{\sf T}}{2 \alpha }\label{eff3-15}\\
&{\sf T}^{0}_{1}(a)=\frac{{{\sf T}^{0}_{1,0}}^{2}-2 \alpha {\gamma_{1}^{0}} +\sqrt{8 \alpha {\gamma_{1}^{0}}{{\sf T}^{0}_{1,0}}^{2}  a^{12}+\left({{\sf T}^{0}_{1,0}}^{2}-2 \alpha{\gamma_{1}^{0}}\right)^{2}}}{2{\sf T}^{0}_{1,0} a^6},\label{eff3-16}\\
&\rho_{\rm{(eff)},1}^{0}({\sf T})=\frac{\alpha\gamma_{1}^{0}}{2 {\sf T}}-\frac{{\sf T}}{4}\label{eff3-17}.
\end{align}
The last example for this subsection, is the appearance of an effective de Sitter state. In the case of $\mathcal{W}_{\rm{(eff)}}=-1$, the following results can be obtained
\begin{align}
h^{w}_{-1}({\sf T})=-\frac{{\sf T}}{\alpha}+\gamma_{-1}^{w}.\label{eff3-18}
\end{align}
For the above function two possibilities could occur. For $w\neq1$ we have
\begin{align}
&{\sf T}^{w\neq1}_{-1}={\sf T}^{w\neq1}_{-1,0},\label{eff3-19}\\
&\rho_{\rm{(eff)},-1}^{w\neq1}=\frac{\alpha  \gamma_{-1}^{w\neq1}(1-3 w)+{\sf T}^{w\neq1}_{-1,0}(w-1)}{6 w-2}\label{eff3-20}.
\end{align}
Therefore, a perfect fluid with arbitrary value of $w$ always leads to a de Sitter phase at late times provided that, the function (\ref{eff3-18}) is chosen. For the case of stiff fluid equation (\ref{relation-w}) reduces to the standard conservation equation when (\ref{eff3-18}) is chosen, therefore, in this case we have
\begin{align}
&{\sf T}^{1}_{-1}={\sf T}^{1}_{-1,0}a^{-6},\label{eff3-19}\\
&\rho_{\rm{(eff)},-1}^{1}=-\frac{\alpha  \gamma_{-1}^{1}}{2}\label{eff3-20}.
\end{align}
This result coincides with the results of subsection (\ref{sub1}) for the stiff fluid with constant effective density.
\subsection{Effective fluid which satisfies, $d\rho_{(\rm{eff})}/d{\sf T}=\mathcal{G}({\sf T},\rho_{(\rm{eff})},p_{(\rm{eff})})$}\label{sub4}
\begin{figure}[h!]
\centering
\centerline{\epsfig{figure=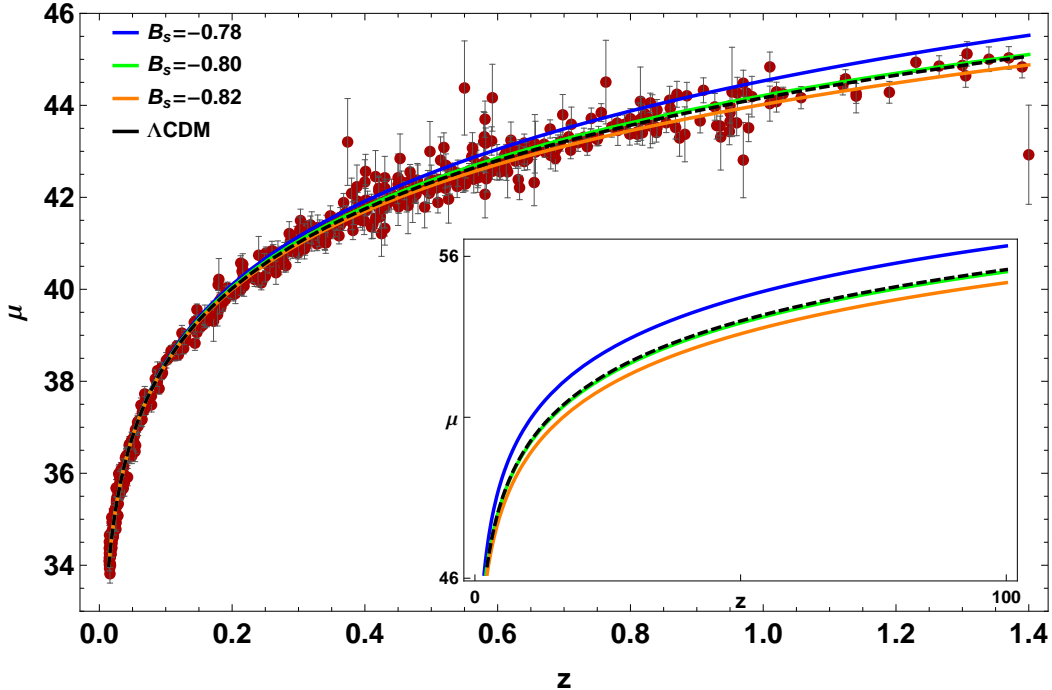,width=14cm}}
\caption{The modulus distance for an effective fluid which behaves as a generalized Chaplygin gas with $p=A/\rho^{1/2}$ for three different values of $B_{s}$. The black dashed line shows the corresponding plot for the $\Lambda${\sf CDM} model. The purple dots with their error bars show the Union 2 compilation data.}
\label{fig2}
\end{figure}

\begin{figure}[h!]
\centering
\centerline{\epsfig{figure=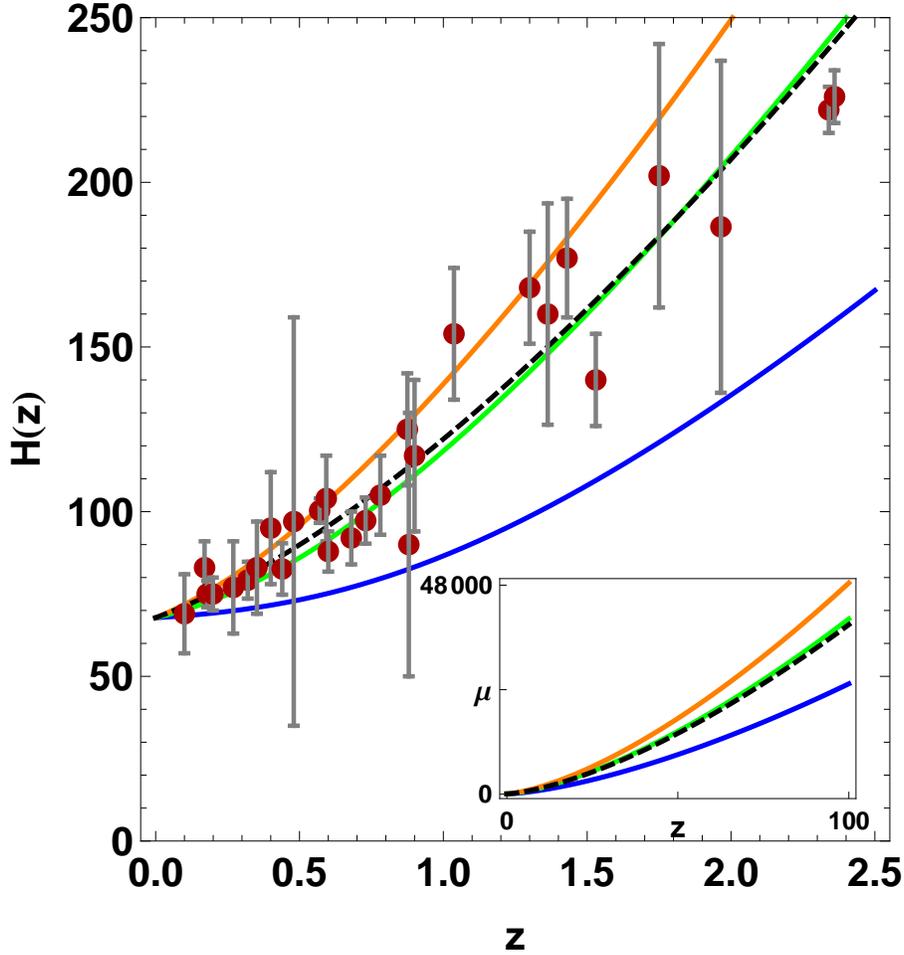,width=12cm}}
\caption{The modulus distance for an effective fluid which behaves as a generalized Chaplygin gas with $p=A/\rho^{1/2}$ for the same values of $B_{s}$ as Figure~\ref{fig2}. The observational data have been used from Farooq, et al.~\cite{farooqetal}}
\label{fig3}
\end{figure}

In this subsection, we consider models for which one can write variation of the effective density with respect to trace in terms of an arbitrary function of the trace, the effective density and the effective pressure. For such models conservation equation (\ref{re7}) takes the following form
\begin{align}\label{eff3-20}
\mathcal{G}({\sf T},\rho_{(\rm{eff})},p_{(\rm{eff})})\frac{d{\sf T}}{da}+\frac{3}{a}(\rho_{(\rm{eff})}+p_{(\rm{eff})})=0.
\end{align}
Equation (\ref{eff3-20}) can be generally a complicated differential equation, however, we consider the most simple examples. As the first one, we study models with $d\rho_{(\rm{eff})}/d{\sf T}=[n/(1+w){\sf T}](\rho_{(\rm{eff})}+p_{(\rm{eff})})$,  where $n$ is a constant. Substituting for the effective quantities from definitions (\ref{re2}) and (\ref{re6}) we get
\begin{align}\label{eff3-21}
2\alpha {\sf T} (w+1) h''({\sf T})-\alpha  (2 n+w-3) h'({\sf T})-2(n-1)=0,
\end{align}
with the solution
\begin{align}\label{eff3-22}
h({\sf T})=\frac{2 \mathcal{D} (w+1)}{2 n+3 w-1} {\sf T}^{\frac{3}{2}+\frac{n-2}{w+1}}-\frac{2 (n-1)}{\alpha  (2 n+w-3)}{\sf T},
\end{align}
where $\mathcal{D}$ is an integration constant. On the other hand, equation (\ref{eff3-20}) takes the following simple form 
\begin{align}\label{eff3-23}
\frac{d{\sf T}}{da}+\frac{3n(1+w)}{a}{\sf T}=0,
\end{align}
which has the solution
\begin{align}\label{eff3-24}
{\sf T}={\sf T}_{0}a^{-\frac{3 (w+1)}{n}}.
\end{align}
As can be seen, in the case of $n=1$ we find a class of models which respects the conservation of {\sf EMT} in $f({\sf R},{\sf T})$ gravity. These solutions have been discussed in several papers. Note that, one can eliminate ${\sf T}$ from (\ref{re2}) and (\ref{re6}) to obtain a relation as $p_{(\rm{eff})}=\mathcal{Q}(\rho_{(\rm{eff})})$, for instance in the well-known dust case with $w=0$ (for which the conservation of {\sf EMT} is automatically met in $f({\sf R},{\sf T})$ gravity) we get
\begin{align}\label{eff3-244}
p_{\text{eff}}= \alpha^2 \mathcal{D}^2 \left(\pm 1 \pm \sqrt{1-\frac{\rho _{\text{eff}}}{\alpha ^2 \mathcal{D}^2 }}\right).
\end{align}
It means that in minimal $f({\sf R},{\sf T})$ gravity a dust perfect fluid behaves as an effective fluid with a non-linear {\sf EoS}. In the next example we consider a class of models for which we can write $\mathcal{G}({\sf T},\rho_{(\rm{eff})},p_{(\rm{eff})})=m$ where $m$ being a constant parameter. In this case we obtain 
\begin{align}\label{eff3-25}
2\alpha (w+1){\sf T} h''({\sf T})-\alpha  (w-3) h'({\sf T})+2[1+m(1-3w)]=0,
\end{align}
with the following solution
\begin{align}\label{eff3-26}
h({\sf T})=\frac{2\mathcal{E} (w+1)}{3 w-1} {\sf T}^{\frac{3}{2}-\frac{2}{w+1}}+\frac{2[1+m(1-3 w)]}{\alpha  (w-3)}{\sf T},
\end{align}
where a constant term has been dropped. Substituting function ($\ref{eff3-26}$) into definitions given for the effective density and pressure we arrive at the following expressions for effective energy density and pressure profile as
\begin{align}
&\rho _{(\rm{eff})}=m {\sf T}=m(3w-1)\rho,\label{eff3-27}\\
&p _{(\rm{eff})}=\frac{w^2-m(1-3 w)^2-1}{(w-3) (3 w-1)}{\sf T}+\frac{\alpha  \mathcal{E} (w+1)}{3 w-1}{\sf T}^{\frac{3}{2}-\frac{2}{w+1}}.\label{eff3-28}
\end{align}
Eliminating trace ${\sf T}$ from ($\ref{eff3-27}$) and ($\ref{eff3-28}$) leaves us with a familiar {\sf EoS}, i.e., 
\begin{align}\label{eff3-29}
p _{(\rm{eff})}=B\rho _{(\rm{eff})}-\frac{A}{\rho _{(\rm{eff})}^{\gamma}},
\end{align}
where
\begin{align}
&B\equiv\frac{-m(1-3 w)^2+w^2-1}{m(w-3) (3 w-1)},\label{eff3-30}\\
&A\equiv\frac{\alpha  \mathcal{E} (w+1)}{3 w-1}m^{\gamma},\label{eff3-31}\\
&\gamma\equiv -\frac{3}{2}+\frac{2}{w+1}.\label{eff3-32}
\end{align}
Therefore, using conservation equation (\ref{re7}) for the effective fluid we obtain the form of effective density as
\begin{align}\label{eff3-33}
\rho _{(\rm{eff})}=\rho _{(\rm{eff}),0}\left[B_{s}+(1-B_{s})^{-3(1+B)(1+\gamma)}\right]^{\frac{1}{1+\gamma}},
\end{align}
where $B_{s}\equiv A/(1+B)\rho _{(\rm{eff}),0}^{1+\gamma}$. The above results show that a perfect fluid with arbitrary value of {\sf EoS} parameter (except for those which make equations (\ref{eff3-30})-(\ref{eff3-32}) singular) could effectively behave as a modified Chaplygin gas~\cite{Benaoum02} provided that function (\ref{eff3-26}) is chosen. For the particular case $w=0$ and $m=-1$ (which has been already discussed in~\cite{Shabani2017-2}) the coefficient $B$ becomes null and in this case a pressure-less matter behaves as a generalized Chaplygin gas with $\gamma=1/2$. We do not intend to confront these models to the observationally data, since, these investigations invoke a comprehensive study. However, we have plotted the modulus distance and the Hubble parameter as well as their observational data for the case with $w=0, m=-1$ in Figures~\ref{fig2} and~\ref{fig3}, respectively. Figure~\ref{fig2} shows an excellent compatibility with the $\Lambda${\sf CDM} model even in the high red-shift regimes. In  Figure~\ref{fig3} there can be seen a slightly deviation from the $\Lambda${\sf CDM} model at low redshifts, however, this deviation remains negligible in high redshift. Though the aforementioned models can be utilized for cosmological settings, a thorough quantified comparison must be performed to investigate the viability of these models.
\section{Concluding remarks}\label{con}
In this work we studied $f({\sf R},{\sf T})={\sf R}+h({\sf T})$ cosmological models from a novel point of view including a single perfect fluid where ${\sf T}$ denotes the trace of {\sf EMT}. A direct approach to analyze these type of models is to choose the functionality of $h({\sf T})$ and then substitute it back into the equation which can be obtained through applying the Bianchi identity on the field equations. Therefore, in this way, we can in principle find the trace in terms of the scale factor. Finally, we can proceed with solving the Friedman equations in order to investigate the cosmological consequences of employing such an $h({\sf T})$ function. In this routine strategy there are at least two difficulties or ambiguities. Firstly, there generally is no guideline to choose a suitable form for $h({\sf T})$ function. It can be more intuitive if we could find the function $h({\sf T})$ based upon a physical principle. Another issue is that the behavior of the matter is not clear. There only is a function $h({\sf T})$ as an input assumption and then achieving a cosmological scenario as the output. However, it is not obvious that how matter behaves to eventuate a physically reasonable cosmological output. The latter problem is more important for theories in which matter or a feature of matter like its energy-momentum trace couples to geometry in an unusual way. Thus, a question that may come to mind is that, what can be the effective behavior of ordinary matter in the context of e.g., $f({\sf R},{\sf T})$ gravity. This paper is a preliminary attempt in order to address the above issues.

In order to give a better understanding of our idea, we reformulated the minimal coupling models $f({\sf R},{\sf T})=R+h({\sf T})$ in terms of the components of an effective fluid. In this approach it is implicitly assumed that the actual matter content of the Universe (as described by a perfect fluid) have an effective behavior which can be described or translated in terms of the density and pressure of a sort of fluid which we have called it ``the effective fluid". Moreover, we have shown that the {\sf EMT} of this effective fluid is automatically conserved. In this regard, the above two problems will be resolved in such a way that for a particular property of the effective fluid, the form of $h({\sf T})$ function will be determined. For example, we can construct a class of (minimal) $f({\sf R},{\sf T})$ cosmological models for which the effective density or pressure are constant, or even the models that accept a relation between the effective density and pressure profile. Therefore, in terms of the effective fluid language, we are able to find $h({\sf T})$ functions based on some physical concepts. On the other hand, we can describe or translate the behavior of the actual perfect fluid (which acts as the seed for either producing or interacting with geometry) in terms of an effective fluid. In this way, the overall effects of the perfect fluid (which originates from an unusual contribution within the Friedmann equations) will be encoded into an effective fluid through the standard Friedmann equations. 

It is informative to point out that the thermodynamic interpretation of models with geometry-matter coupling has been beautifully discussed by Harko in~\cite{Harko2014} where, he has studied the modified conservation equation which appears in these theories using the formalism of open thermodynamic systems. The author has concluded that such a conservation equation corresponds to an irreversible matter creation process which in turn implies the space-time transmutation into matter. Nevertheless, in terms of the effective fluid, there only is a conservative energy flow between the effective fluid and geometry without the need of particle creation from space-time.

Specifically, in terms of the effective fluid we have considered four different class of models. The first class of models are those for which the effective density is constant. We have shown that this assumption corresponds to a de Sitter state at late times. Only the stiff fluid with $w=1$ and the {\sf DE} with $w=-1$ satisfies such an assumption. We have also considered a class of models with constant effective pressure. We illustrated that these models accept a transition between a dust dominated to an accelerated expansion era. We have discussed that in these models the necessity for including a cosmological constant can be translated into the pressure of the effective fluid. The third class of models are those for which a perfect fluid with $w$ has the behavior of an effective perfect fluid with $\mathcal{W}_{\rm{(eff)}}$. In this category, different possibilities have been considered. As the last example, we have investigated two different class of models in which we have $\mathcal{G}({\sf T},\rho_{(\rm{eff})},p_{(\rm{eff})})dT/da+\frac{3}{a}(\rho_{(\rm{eff})}+p_{(\rm{eff})})=0$ where $\mathcal{G}({\sf T},\rho_{(\rm{eff})},p_{(\rm{eff})})=d\rho_{(\rm{eff})}/d{\sf T}$. Furthermore, we have shown that the models for which the condition $d\rho_{(\rm{eff})}/dT=[n/(1+w){\sf T}](\rho_{(\rm{eff})}+p_{(\rm{eff})})$ holds, admit solution ${\sf T}={\sf T}_{0}a^{-3(1+w)/n}$. These models which make the {\sf EMT} of the actual perfect fluid conserved  have been frequently discussed in the literature for $n=1$. In the case of $w=0$, one can depict the overall behavior of matter in terms of the behavior of an effective fluid with pressure given by $p_{\text{eff}}= \alpha^2 \mathcal{D}^2 \left(\pm 1 \pm \sqrt{1-\rho _{\text{eff}}/(\alpha ^2 \mathcal{D}^2 )}\right)$. As the last example we considered models for which we have $d\rho_{(\rm{eff})}/dT=m$ and we showed that these type of models effectively behave like a modified Chaplygin gas with four free parameters.

Finally, we note that the aim of this work has been to bring forward an alternative view to $f({\sf R},{\sf T})$ gravity using the idea of an effective cosmic fluid. However, the study of cosmological consequences of these models demand a comprehensive quantified study and we did not consider such an investigation within the present article. Nevertheless, work on these issues can for example be pursued by considering the modified Chaplygin gas representation of these models that have been discussed in the present work, using different cosmological tests.

\end{document}